\documentstyle[preprint,prc,aps,psfig]{revtex}

\renewcommand{\vec}[1]{\mbox{\boldmath $#1$}}

\tightenlines

\begin{document}
\title{Radiation correction to astrophysical fusion reactions 
and the electron screening problem} 
\author{K. Hagino$^1$ and A.B. Balantekin$^{2,3}$}
\address{$^1$Yukawa Institute for Theoretical Physics, Kyoto
University, Kyoto 606-8502, Japan }
\address{$^2$Department of Physics, University of Wisconsin, 
Madison, WI 53706, USA}
\address{$^3$European Centre for Theoretical Studies in Nuclear
Physics and Related Areas (ECT*), Villa Tambosi, I-38050, Villazzano,
Trento, Italy}

\maketitle

\begin{abstract}

We discuss the effect of electromagnetic environment 
on laboratory measurements of the 
nuclear fusion reactions of astrophysical interest. 
The radiation field is eliminated using the path integral 
formalism in order to obtain the influence functional, which we 
evaluate in the semi-classical approximation. 
We show that enhancement of the tunneling probability due to the 
radiation correction is extremely small and 
does not resolve the longstanding problem that the observed electron 
screening effect is significantly larger than theoretical predictions. 
\end{abstract}

\pacs{PACS numbers: 25.10.+s,03.65.Sq, 12.20.-m, 24.90.+d}

Nuclear fusion reactions measured at a laboratory at very low 
incident energies 
are subjected to the electron screening effect, which originates from bound
electrons in the target atom (or molecule). 
The electrons shield the Coulomb potential between the projectile and
the target nuclei, and it is expected that fusion cross sections are 
enhanced over the prediction which does not take into 
account this correction \cite{ALR87}. Indeed, since the seminal work for the
$^3$He(d,p)$^4$He reaction \cite{EKN88}, the experimental fusion cross
sections (or equivalently the astrophysical S factors) are shown to be 
enhanced as compared to those extrapolated from the high energy 
region where the electron screening effect is negligible down to the 
low energy regime (see Ref. \cite{SRSC01} for a recent review).  

In the low energy limit, the adiabatic approximation is validated, and 
the electron screening effect can be expressed as a constant energy
shift of the Coulomb potential, where the amount of the shift is given
by a difference of electron binding energies between the unified and the
isolated systems \cite{ALR87}. The experimental data, however,
systematically indicate that a significantly larger energy shift is
required in order to account for the low energy enhancement of the 
fusion cross sections \cite{L93}. This has been a big surprise and 
also a rather puzzling situation, 
since the
adiabatic approximation should provide the upper limit of the tunneling
probability and thus the upper limit of the energy shift
\cite{THAB94,THA95,SKLS93}. No satisfactory explanation has been
proposed to reconcile this problem so far. Furthermore, the recent careful
measurement for the stopping power \cite{ARG01,CFJ00} as well as the
recent attempts which used the Trojan-horse method \cite{B86} to
measure the bare cross sections at low energies \cite{STP01,MPG01} have
re-confirmed that the enhancement of fusion cross section 
is much larger than the model calculation. 

Balantekin {\it et al.} studied several effects beyond the
electron screening on astrophysical reactions, which include vacuum
polarization, relativity, bremsstrahlung, and atomic polarization
\cite{BBH97}. They found that all of these effects are too
small to explain the difference between the measured and the predicted 
electron screening energies. Obviously 
the low energy enhancement of the astrophysical reactions 
has a different origin. 
Recently, Flambaum and Zelevinsky suggested that the
virtual photon emission during tunneling may increase the
penetrability \cite{FZ99}. Within a few approximations, which include 
the closure approximation for the relative motion between the
projectile and the target nuclei, they derived 
a static potential shift due to the radiation correction which is 
proportional to the second derivative of the bare internucleus
potential. The 
second derivative is exactly zero for the point Coulomb potential and 
the potential renormalization which Flambaum and Zelevinsky derived 
has a finite value only well inside the Coulomb potential. 
Note that such
potential renormalization can be well compensated with a choice of a
nuclear potential between the projectile and the target unless the energy
dependence is very strong, as in the case of coupling
to high-lying states in heavy-ion subbarrier fusion reactions
\cite{HTDHL97}, and it is thus not easy to separate out the radiation 
effect in a clear way. 
Since the closure approximation may be too crude for astrophysical
reactions, it is necessary to re-examine the effect of virtual photon
emission without resorting to the approximations which Flambaum and
Zelevinsky used in order to assess the role of radiation field in
realistic systems. 

The purpose of this paper is to study the effect of coupling between
the tunneling motion and the electromagnetic field taking a different
approach from Flambaum and Zelevinsky. To this end, we employ the 
path-integral formalism for multi-dimensional tunneling \cite{BT85}
and evaluate the influence functional in the semi-classical
approximation. Besides the semi-classical and the dipole
approximations, our formalism
is exact, and the effect is finite even with the pure Coulomb
potential. The path integral approach allows us to discuss both 
virtual and real photon emissions during
tunneling on the equal footing. In this respect,
our study is relevant also to the bremsstrahlung in $\alpha$
decay, which has attracted much interest for the past few years
\cite{K97,PB98,TNH99}.  

Consider the tunneling motion in the presence of the
radiation field. 
The classical Lagrangian for the system reads
\cite{F50,FH65} 
\begin{eqnarray}
L&=&L_0 + L_{\rm em} + L_{\rm int}, \label{lagrangian} \\
L_0&=&\frac{\mu}{2}\dot{\vec{r}}^2-V(r), \label{lagrangian0}\\
L_{\rm em}&=&
\int\frac{d^3k}{(2\pi)^3}\sum_\alpha 
\left(\frac{1}{2}\dot{Q}_{\vec{k}\alpha}^2-\frac{1}{2}
\omega^2 Q_{\vec{k}\alpha}^2\right), \label{lagrangian1}\\
L_{\rm int}&=&\sqrt{4\pi}e_{\rm E1}\int\frac{d^3k}{(2\pi)^3}\sum_\alpha 
(\dot{\vec{r}}\cdot\vec{\epsilon}_{\vec{k}}^{(\alpha)})\,
Q_{\vec{k}\alpha}, \label{lagrangian2}
\end{eqnarray}
where $\alpha$ represents the polarization index and
$\vec{\epsilon}_{\vec{k}}^{(\alpha)}$ is the corresponding
polarization vector. We employ the Coulomb gauge, and thus the
polarization vectors are orthogonal to the photon momentum $\vec{k}$. 
We have used the dipole approximation, and introduced the photon
coordinate $Q_{\vec{k}\alpha}(t)$, which 
is independent of the relative motion $\vec{r}$ between the 
projectile and the target nuclei. $e_{\rm E1}$ is the E1 effective
charge given by $(A_PZ_T-A_TZ_P)/(A_P+A_T)$, $A_T$ and $Z_T$ being 
the mass and the atomic numbers for the target nucleus, respectively, 
and similar for the projectile also. $\mu$ is the reduced mass for the 
relative motion $\vec{r}$, and $V(r)$ is the potential between the
projectile and the target in the absence of the radiation field. 
As usual, we have neglected a term which is proportional to 
$e_{\rm E1}^2$ in the classical Lagrangian. 

Our interest is to compute the transition amplitude for the relative
motion from $\vec{r}_i$ to $\vec{r}_f$ while for the radiation field
from the vacuum state $|0\rangle$ to the $n$-photon state
$|n\rangle$ for a given incident energy $E$. This is expressed in 
terms of the energy representation of the path integral amplitude as 
\begin{equation}
{\cal T}_n=g\int_0^\infty dT\,e^{iET/\hbar}
\int{\cal D}[\vec{r}(t)] 
e^{(i/\hbar)\int_0^Tdt\,L_0} 
\left\langle n\left|
\int{\cal D}[Q_{\vec{k}\alpha}(t)] 
e^{(i/\hbar)\int_0^Tdt\,(L_{\rm em}+L_{\rm int})} \right|0\right\rangle,
\end{equation}
where the kinematical factor $g$ is proportional to $\sqrt{E/\mu}$ \cite{BT85}. 
Here the integral for $\vec{r}$ is carried out for all paths which 
connect $\vec{r}(0)=\vec{r}_i$ and $\vec{r}(T)=\vec{r}_f$. 
We square this amplitude and sum over $n$ in order to obtain the total 
transition probability for the relative motion: 
\begin{eqnarray}
P&=&\sum_n|{\cal T}_n|^2 \\
&=&|g|^2\int_0^\infty dT\,e^{iET/\hbar}
\int_0^\infty d\widetilde{T}\,e^{-iE\widetilde{T}/\hbar} \nonumber \\
&&\times 
\int{\cal D}[\vec{r}(t)] 
\int{\cal D}[\tilde{\vec{r}}(\tilde{t})] 
e^{(i/\hbar)\int_0^Tdt\,L_0(\vec{r})} 
e^{-(i/\hbar)\int_0^{\widetilde{T}}d\tilde{t}\,L_0(\tilde{\vec{r}})}
\rho(\tilde{\vec{r}}(\tilde{t}),\widetilde{T};\vec{r},T), 
\label{penetrability}
\end{eqnarray}
where $\rho$ is the two-time influence functional given by \cite{BT85}
\begin{eqnarray}
\rho(\tilde{\vec{r}}(\tilde{t}),\widetilde{T};\vec{r},T)
&=&
\left\langle 0\left|
\int{\cal D}[\tilde{Q}_{\vec{k}\alpha}(\tilde{t})] 
e^{-(i/\hbar)\int_0^{\widetilde{T}}d\tilde{t}\,
(L_{\rm em}(\tilde{Q})+L_{\rm int}(\tilde{\vec{r}},\tilde{Q}))} 
\right.\right.
\nonumber \\
&& \times 
\left.\left.
\int{\cal D}[Q_{\vec{k}\alpha}(t)] 
e^{(i/\hbar)\int_0^T dt\,
(L_{\rm em}(Q)+L_{\rm int}(\vec{r},Q))} 
\right|0\right\rangle.
\end{eqnarray}
For the Lagrangian given in Eqs. (\ref{lagrangian}) - (\ref{lagrangian2}), 
the two-time influence functional can be expressed analytically and is
given by \cite{BT85,F50,FH65}
\begin{eqnarray}
&&\rho(\tilde{\vec{r}}(\tilde{t}),\widetilde{T};\vec{r},T)
=
\exp\left[
\int\frac{d^3k}{(2\pi)^3}\sum_\alpha \left\{
-\frac{i}{2}\omega(T-\widetilde{T})
-\frac{1}{2\hbar\omega}
\left(
\int^T_0dt\int^t_0ds\,
f_{k\alpha}(t)f_{k\alpha}(s)e^{-i\omega(t-s)}
\right.\right.\right. \nonumber \\
&&
\quad\quad\left.\left.\left.
+\int^{\widetilde{T}}_0dt\int^t_0ds\,
\tilde{f}_{k\alpha}(t)\tilde{f}_{k\alpha}
(s)e^{i\omega(t-s)}
-e^{-i\omega(T-\widetilde{T})}
\int^T_0dt\,f_{k\alpha}(t)e^{i\omega t}
\int^{\widetilde{T}}_0ds\,\tilde{f}_{k\alpha}(s)e^{-i \omega s}
\right)\right\}\right]
\label{influence}
\end{eqnarray}
where $f_{k\alpha}$ and $\tilde{f}_{k\alpha}$ is defined by 
\begin{equation}
f_{k\alpha}(t)\equiv -\sqrt{4\pi}e_{\rm E1}\dot{\vec{r}}\cdot
\vec{\epsilon}_{\vec{k}}^{(\alpha)},
~~~
\tilde{f}_{k\alpha}(t)\equiv -\sqrt{4\pi}e_{\rm E1}\dot{\tilde{\vec{r}}}
\cdot\vec{\epsilon}_{\vec{k}}^{(\alpha)}.
\end{equation}
In the exponent in the influence functional (\ref{influence}), the 
last term is the real photon contribution, while the 
second and the third terms represent the virtual photon emission
\cite{F50}. 

In the radiation problems such as the Lamb shift calculation, it is
essential to separate out a divergent contribution due to 
the mass renormalization in order to obtain a physical result. 
For our problem, this can be done by performing a partial integration 
for the second and the third terms in Eq.(\ref{influence}):
\begin{eqnarray}
\int^T_0dt\int^t_0ds\,
f_{k\alpha}(t)f_{k\alpha}(s)e^{-i\omega(t-s)}
&=&
\frac{1}{i\omega}\int^T_0dt\,(f_{k\alpha}(t))^2
-\frac{1}{i\omega}\int^T_0dt\,f_{k\alpha}(t)f_{k\alpha}(0)e^{-i\omega
t} \nonumber \\
&&-\frac{1}{i\omega}\int^T_0dt\int^t_0ds\,
f_{k\alpha}(t)\frac{df_{k\alpha}(s)}{ds}e^{-i\omega(t-s)}.
\end{eqnarray}
The first term is proportional to $\dot{\vec{r}}^2$ and is nothing but 
the mass renormalization, which has already included in the
kinetic term in Eq. (\ref{lagrangian0}). The second term vanishes if
we choose $t=0$ at the outer classical turning point (note that we are
going to deal with the tunneling problem in the following). Retaining
only the third term, the regularized influence functional 
$\rho_{\rm reg}$ then reads 
\begin{eqnarray}
&&\rho_{\rm reg}(\tilde{\vec{r}}(\tilde{t}),\widetilde{T};\vec{r},T)
=
\exp\left[
\int\frac{d^3k}{(2\pi)^3}\sum_\alpha 
\left(-\frac{1}{2\hbar\omega}\right)
\left\{-\frac{1}{i\omega}
\int^T_0dt\int^t_0ds\,
f_{k\alpha}(t)\frac{df_{k\alpha}(s)}{ds}e^{-i\omega(t-s)}
\right.\right.\nonumber \\
&&
\left.\left.
+\frac{1}{i\omega}\int^{\widetilde{T}}_0dt\int^t_0ds\,
\tilde{f}_{k\alpha}(t)\frac{d\tilde{f}_{k\alpha}(s)}{ds}e^{i\omega(t-s)}
-e^{-i\omega(T-\widetilde{T})}
\int^T_0dt\,f_{k\alpha}(t)e^{i\omega t}
\int^{\widetilde{T}}_0ds\,\tilde{f}_{k\alpha}(s)e^{-i \omega s}
\right\}\right].
\end{eqnarray}
Here, we have dropped also the $-\frac{i}{2}\omega(T-\widetilde{T})$
term, since it merely changes the definition of the energy of the
vacuum state. 

We now introduce the semi-classical approximation to the path integrals
with respect to $\vec{r}$ as well as to the time integrals in
Eq. (\ref{penetrability}) \cite{THA95,TB84,LT83}. This leads to a
classical tunneling path from the outer to the inner turning points 
along the imaginary time axis. For simplicity, in the following, we
consider only a head-on collision. Since the coupling strength 
$e^2_{\rm E1}/\hbar c$ is small, we determine the tunneling path by
disregarding the radiation field. The classical path thus obeys the
Newtonian equation with the inverted potential, 
\begin{equation}
\mu\frac{d^2z_{\rm cl}}{d\tau^2}=\frac{d}{dz}V(z_{\rm cl}). 
\label{Newton}
\end{equation}
The influence functional is then evaluated along the classical path. 
After carrying out the angle integral for the photon momentum
$\vec{k}$, the penetrability (\ref{penetrability}) in the
semi-classical approximation thus reads 
\begin{equation}
P=P_0\cdot f,
\end{equation}
where 
\begin{eqnarray}
P_0
&=&|g|^2\int_0^\infty dT\,e^{iET/\hbar}
\int_0^\infty d\widetilde{T}\,e^{-iE\widetilde{T}/\hbar} \nonumber \\
&&\times 
\int{\cal D}[z(t)] 
\int{\cal D}[\tilde{z}(\tilde{t})] 
e^{(i/\hbar)\int_0^Tdt\,L_0(z)} 
e^{-(i/\hbar)\int_0^{\widetilde{T}}d\tilde{t}\,L_0(\tilde{z})},
\end{eqnarray}
is the penetrability in the absence of the radiation field, and 
\begin{eqnarray}
f&=&
\rho_{\rm reg}(z_{\rm cl}(t)^*,T_{\rm cl}^*;z_{\rm cl}(t), T_{\rm cl}) \\
&=&
\exp\left[
\int^\infty_0\frac{k^2dk}{(2\pi)^3}\,
\frac{1}{2\hbar\omega}\,\frac{8\pi}{3}
\left\{\frac{2}{\omega}
\int^{T_0}_0d\tau\int^{\tau}_0d\tau'\,4\pi e_{\rm E1}^2\,
\frac{1}{\mu}\,\frac{dz}{d\tau}\,\frac{dV}{dz(\tau')}
\,e^{-\omega(\tau-\tau')} \right.\right.\nonumber \\
&&
\left.\left.
+4\pi e^2_{\rm E1}\,
e^{-2\omega T_0}\left(\int^{T_0}_0d\tau\,\frac{dz}{d\tau}\,
e^{\omega\tau}\right)^2
\right\}\right].
\label{enhance}
\end{eqnarray}
is the regularized two-time influence functional along the classical
trajectory $z_{\rm cl}$. In obtaining $f$, we have introduced the
imaginary time evolution, $t\to -i\tau, s\to -i\tau'$, and $T\to -i
T_0$, and neglected the excitation of the radiation field prior to the
tunneling. $T_0$ is the Euclidean tunneling time, where 
$z_{\rm cl}(T_0)$ corresponds to the inner turning point 
$R_0$ while $z_{\rm cl}(0)$ is the outer turning point $Z_PZ_Te^2/2E$. 

Let us now numerically estimate the enhancement factor $f=P/P_0$ for
the $d$+$^3$He reaction. To this end, we consider the pure
Coulomb potential $V(r)=Z_PZ_Te^2/r$ from the outer turning point 
$Z_PZ_Te^2/E$ to the inner turning point $R_0$=4.3 fm. 
The Newtonian equation (\ref{Newton}) can then 
be solved analytically as, 
\begin{equation}
\tau=\frac{Z_PZ_Te^2}{2E}\sqrt{\frac{\mu}{2E}}(w+\sin w), ~~~~~~~
z(\tau)=
\frac{Z_PZ_Te^2}{2E}(1+\cos w).
\end{equation}
Because of the exponential factor, the $k$ integral in
Eq. (\ref{enhance}) is quickly damped as a function of $k$, and it is not
necessary to introduce the momentum cut-off factor for our tunneling problem. 
Figure 1 shows the deviation of the enhancement factor $f$ from unity 
as a function of the incident energy. We see that the enhancement due to
the radiation coupling in the tunneling region is extremely small and
does not play any important role in the astrophysical fusion reaction. 
The dashed line shows the virtual photon contribution separately,
which is obtained by neglecting the second term in
Eq. (\ref{enhance}). One finds that the virtual photon emission
provides the dominant contribution, although the real photon emission
is not negligible. 
Table 1 shows the enhancement factor for several reactions. We take
the same value for the inner turning point $R_0$ as in
Ref. \cite{BBH97}. In all the reactions considered, the effects of the
radiation coupling during tunneling is almost negligible and the
enhancement factor never exceeds 10$^{-3}$ \%. This effect is even much
smaller than the other small effects considered in Ref. \cite{BBH97}. 
We thus conclude that the radiation correction is not helpful 
for the low energy enhancement of nuclear astrophysical reactions. 

In summary, we have studied the effect of the radiation coupling on
the tunneling motion of charged particles. To this end, we have
employed the semi-classical approximation to the transition amplitude
in the path integral representation. We have shown that the effect is 
almost negligible for nuclear astrophysical reactions, and the large
electron screening problem remains unsolved. 
The radiation correction
which we discussed in this paper is even smaller than the vacuum
polarization effect, and would be negligible as compared to the 
latter. 
The origin of the discrepancy between the measured and the calculated 
electron screening energies still seems to be an open problem. 

\medskip

We thank the European Centre for Theoretical Studies in Nuclear Physics
and Related Areas (ECT*), where a part of this work was done, for its
hospitality and for partial support for this project.
This work was supported in part by the U.S. National Science
Foundation Grant No. PHY-0070161 at the University of
Wisconsin, and in part by the University of Wisconsin Research
Committee with funds granted by the Wisconsin Alumni Research
Foundation.

\begin{figure}
  \begin{center}
    \leavevmode
    \parbox{0.9\textwidth}
           {\psfig{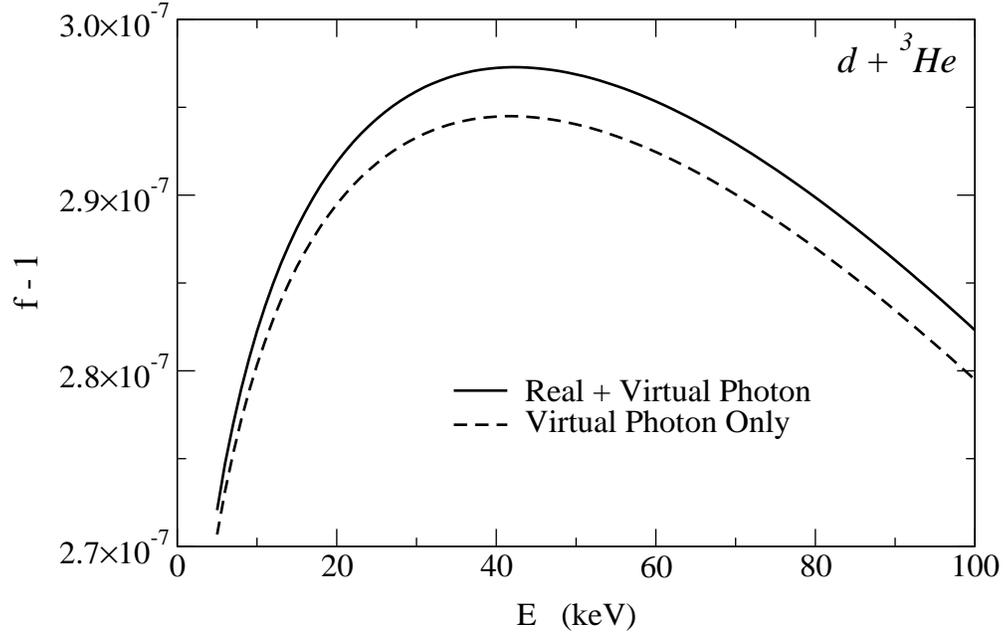}}
  \end{center}
\protect\caption{
The deviation of the enhancement factor $f$ due to the radiation
coupling from unity as a function of the incident energy for the 
$d$+$^3$He fusion reaction. 
The solid line includes
    both the real and the virtual photon emissions during tunneling,
    whereas the dashed line represents only the virtual photon
    contribution. 
}
\end{figure}

\begin{table}[hbt]
\caption{
The enhancement factor $f$ for several reactions obtained at the
lowest experimental energy $E_{\rm min}$ and with the inner turning
point $R_0$ indicated. 
}

\begin{center}
\begin{tabular}{c|c|c|c}
Reaction & $E_{\rm min}$ (keV) & $R_0$ (fm) & $f-1$ \\
\hline
$^3$He(d,p)$^4$He & 5.88 & 4.3 & 2.744 $\times$ 10$^{-7}$ \\
D($^3$He,p)$^4$He & 5.38 & 4.3 & 2.731 $\times$ 10$^{-7}$ \\
$^6$Li(p,$\alpha$)$^3$He & 10.74 & 3.0 & 3.758 $\times$ 10$^{-6}$ \\
$^7$Li(p,$\alpha$)$^4$He & 12.70 & 4.3 & 3.598 $\times$ 10$^{-6}$ \\
H($^6$Li,$\alpha$)$^3$He & 10.94 & 3.0 & 3.761 $\times$ 10$^{-6}$ \\
H($^7$Li,$\alpha$)$^4$He & 12.97 & 4.3 & 3.602 $\times$ 10$^{-6}$ \\
$^{10}$B(p,$\alpha$)$^7$Be & 18.70 & 3.3 & 6.095 $\times$ 10$^{-6}$ \\
$^{11}$B(p,$\alpha$)$^8$Be & 16.70 & 2.0 & 1.169 $\times$ 10$^{-5}$ 

\end{tabular}
\end{center}
\end{table}


\begin{references}

\bibitem{ALR87}H.J. Assenbaum, K. Langanke, and C. Rolfs, 
Z. Phys. {\bf A327}, 461 (1987). 

\bibitem{EKN88}S. Engstler, A. Krauss, K. Neldner, C. Rolfs,
U. Schr\"oder, and K. Langanke, Phys. Lett. B{\bf 202}, 179 (1998). 

\bibitem{SRSC01}F. Strieder, C. Rolfs, C. Spitaleri, and
P. Corvisiero, Natrwissenschaften {\bf 88}, 461 (2001). 

\bibitem{L93}K. Langanke, in {\it Advances in Nuclear Physics}, edited
by J.W. Negele and E. Vogt (Plenum, New York, 1993). 

\bibitem{THAB94}N. Takigawa, K. Hagino, M. Abe, and A.B. Balantekin,
Phys. Rev. C{\bf 49}, 2630 (1994). 

\bibitem{THA95}N. Takigawa, K. Hagino, and M. Abe, Phys. Rev. 
C{\bf 51}, 187 (1995). 

\bibitem{SKLS93}T.D. Shoppa, S.E. Koonin, K. Langanke, and R. Seki, 
Phys. Rev. C{\bf 48}, 837 (1993). 

\bibitem{ARG01}
M. Aliotta, F. Raiola, G. Gy\"urky, A. Formicola,
R. Bonetti, C. Broggini, L. Campajola, P. Corvisiero, H. Costantini,
A. D'Onofrio, Z. F\"ul\"op, G. Gervino, L. Gialanella,
A. Guglielmetti, C. Gustavino, G. Imbriani, M. Junker, P.G. Moroni,
A. Ordine, P. Prati, V. Roca, D. Rogalla, C. Rolfs, M. Romano,
F. Sch\"umann, E. Somorjai, O. Straniero, F. Strieder, F. Terrasi,
H.P. Trautvetter, and S. Zavatarelli, Nucl. Phys. {\bf A690}, 790
(2001). 

\bibitem{CFJ00}
H. Costantini, A. Formicola, M. Junker, 
R. Bonetti, C. Broggini, L. Campajola, P. Corvisiero, 
A. D'Onofrio, A. Fubini, G. Gervino, L. Gialanella,
U. Greife, 
A. Guglielmetti, C. Gustavino, G. Imbriani, 
A. Ordine, P.G. Prada Moroni, P. Prati, V. Roca, D. Rogalla, C. Rolfs, 
M. Romano,
F. Sch\"umann, O. Straniero, F. Strieder, F. Terrasi,
H.P. Trautvetter, and S. Zavatarelli, Phys. Lett. B{\bf 482}, 43
(2000). 

\bibitem{B86}G. Baur, Phys. Lett. B{\bf 178}, 135 (1986). 

\bibitem{STP01}C. Spitaleri, S. Typel, R.G. Pizzone, M. Aliotta,
S. Blagus, M. Bogovac, S. Cherubini, P. Figuera, M. Lattuada,
M. Milin, D. Miljanic, A. Musumarra, M.G. Pellegriti, D. Rendic,
C. Rolfs, S. Romano, N. Soic, A. Tumino, H.H. Wolter, and M. Zadro, 
Phys. Rev. C{\bf 63}, 055801 (2001). 

\bibitem{MPG01}A. Musumarra, R.G. Pizzone, S. Blagus, M. Bogovac,
P. Figuera, M. Lattuada, M. Milin, D. Miljanic, M.G. Pellegriti,
D. Rendic, C. Rolfs, N. Soic, C. Spitaleri, S. Typel, H.H. Wolter, and
M. Zadro, Phys. Rev. C{\bf 64}, 068801 (2001). 

\bibitem{BBH97}A.B. Balantekin, C.A. Bertulani, M.S. Hussein,
Nucl. Phys. {\bf A627}, 324 (1997). 

\bibitem{FZ99}V.V. Flambaum and V.G. Zelevinsky, Phys. Rev. Lett. {\bf
83}, 3108 (1999). 

\bibitem{HTDHL97}K. Hagino, N. Takigawa, M. Dasgupta, D.J. Hinde, and
J.R. Leigh, Phys. Rev. Lett. {\bf 79}, 2014 (1997). 

\bibitem{BT85}A.B. Balantekin and N. Takigawa, Ann. of Phys. (N.Y.)
{\bf 160}, 441 (1985); A.B. Balantekin and N. Takigawa,
Rev. Mod. Phys. {\bf 70}, 77 (1998). 

\bibitem{K97}J. Kasagi, H. Yamazaki, N. Kasajima, T. Ohtsuki, and
H. Yuki, Phys. Rev. Lett. {\bf 79}, 371 (1997). 

\bibitem{PB98}T. Papenbrock and G.F. Bertsch, Phys. Rev. Lett. {\bf
80}, 4141 (1998). 

\bibitem{TNH99}N. Takigawa, Y. Nozawa, K. Hagino, A. Ono, and
D.M. Brink, Phys. Rev. C{\bf 59}, R593 (1999). 

\bibitem{F50}R.P. Feynman, Phys. Rev. {\bf 80}, 440 (1950). 

\bibitem{FH65}R.P. Feynman and A.R. Hibbs, {\it Quantum Mechanics and
Path Integrals} (McGraw-Hill, New York, 1965), Chapter 9. 

\bibitem{TB84}N. Takigawa and G.F. Bertsch, Phys. Rev. C{\bf 29},
2358 (1984). 

\bibitem{LT83}S.Y. Lee and N. Takigawa, Phys. Rev. C{\bf 28}, 1123
(1983). 

\end{references}
\end{document}